\documentclass[a4paper]{article}

\usepackage{INTERSPEECH2021}
\usepackage{float}
\usepackage{url}

\renewcommand{\paragraph}[1]{\noindent\newline\textbf{#1}\quad}

\title{Label-Efficient Self-Supervised Speaker Verification\\With Information Maximization and Contrastive Learning}

\name{Theo Lepage, Reda Dehak}

\address{
  Speaker and Language Recognition Group (ESLR), \\
  Laboratoire de Recherche de l'EPITA, France
}

\email{\{theo.lepage, reda.dehak\}@epita.fr}

\begin{document}

\abovedisplayskip=8pt
\abovedisplayshortskip=0pt
\belowdisplayskip=8pt
\belowdisplayshortskip=0pt
\setlength{\parskip}{0cm}

\maketitle

\begin{abstract}
    State-of-the-art speaker verification systems are inherently dependent on some kind of human supervision as they are trained on massive amounts of labeled data. However, manually annotating utterances is slow, expensive and not scalable to the amount of data available today. In this study, we explore self-supervised learning for speaker verification by learning representations directly from raw audio. The objective is to produce robust speaker embeddings that have small intra-speaker and large inter-speaker variance. Our approach is based on recent information maximization learning frameworks and an intensive data augmentation pre-processing step. We evaluate the ability of these methods to work without contrastive samples before showing that they achieve better performance when combined with a contrastive loss. Furthermore, we conduct experiments to show that our method reaches competitive results compared to existing techniques and can get better performances compared to a supervised baseline when fine-tuned with a small portion of labeled data.\\
\end{abstract}

\noindent\textbf{Index Terms}: Speaker Recognition, Self-Supervised Learning, Speaker Embeddings, Label-efficient.

\section{Introduction}

\let\thefootnote\relax\footnotetext{The code associated with this article is publicly available at\\ \url{https://github.com/theolepage/sslsv}.}

Speaker Recognition (SR) is the process of recognizing the identity of the person speaking on an input speech audio. One of its common applications is Speaker Verification (SV) that aims to verify whether a given audio sample matches the identity claimed by a speaker. Recent SR methods are based on learning a speaker embedding space named i-vectors \cite{dehakivector2011} or x-vectors \cite{Snyder2018Xvectors} which try to minimize the intra-speaker and maximize the inter-speaker variabilities. In the last decade, there has been a surge of interest in applying neural networks to SR \cite{villalba2020}. First DNN were used in combination with the i-vector method or as a bottleneck features extractor. Recently DNN methods compute directly a sequence embedding based solely on deep networks. The first DNN method which surpasses the i-vector performance is the x-vector method \cite{Snyder2018Xvectors}. Several variants of DNN x-vectors methods \cite{Chung2020Delving, Chung2020Metric} are used in SR to model speaker characteristics using an embedding space. They are studying alternative network architectures, pooling methods and training objectives with the aim of reaching better performance. All these methods are trained in a supervised way with large labeled datasets. Although impressive progress has been made with supervised learning, this paradigm is now considered as bottleneck for building more intelligent systems. Manually annotating data is a complex, expensive and tedious process, especially when dealing with signals such as images, text and speech. Moreover, the risk is to create biased models that do not work well in real life notably in difficult acoustic conditions.\\

Over the past few years, the concept of self-supervised learning has proven to be very effective to benefit from large amounts of unlabeled data. Self-supervised learning aims at learning meaningful information directly from the data itself without any human supervision. Most methods are considered \textit{contrastive} as their objective is to maximize the similarity between representations extracted from the same temporal context. When dealing with audio samples, the assumption is that segments extracted from the same utterance belong to the same speaker but those from different utterances belong to different speakers. This assumption does not always hold (``class collision'' problem) but the impact on the training convergence is considered negligible. Following this principle, most contrastive learning frameworks, such as SimCLR \cite{Zhang2021SimCLR} and MoCo \cite{Xia2021MOCOSV}, have been successfully applied to the speaker verification task. These methods are based on InfoNCE loss which forces distances of positive pairs to be small and distances of negative pairs to be large in a latent space \cite{Oord2019Representation, Chen2020SimCLR, He2020MOCO}.\\

Contrastive methods are a reliable approach to avoid the \textit{collapse} problem appearing when the model produces constant and non-informative representations. However, there remain many challenges for contrastive self-supervised systems as they rely on a costly negative sampling step that requires large batch sizes or memory banks. Moreover, other techniques avoid a \textit{collapse} by exploiting complex architecture tricks or by adding constraints on the architecture: batch-wise or feature-wise normalization, Siamese architectures with shared weights (SimCLR \cite{Chen2020SimCLR}), a stop gradient operation (SimSiam \cite{chen2020SimSiam}), a ``momentum encoder'' (BYOL \cite{Grill2020BYOL}), vector quantization (SwAV \cite{Caron2021SwAV}) or clustering (DeepCluster \cite{Caron2018DeepCluster}).  Nonetheless, some methods, such as VICReg \cite{Bardes2021VICReg} and Baylow Twins \cite{Zbontar2021Barlow}, were recently introduced to avoid collapse by explicitly maximizing the information from the embeddings. These techniques are simpler and do not require contrastive samples, which open a new range of possibilities regarding speaker verification models trained in a self-supervised fashion.\\

\begin{figure*}[t]
  \centering
  \includegraphics[width=\linewidth]{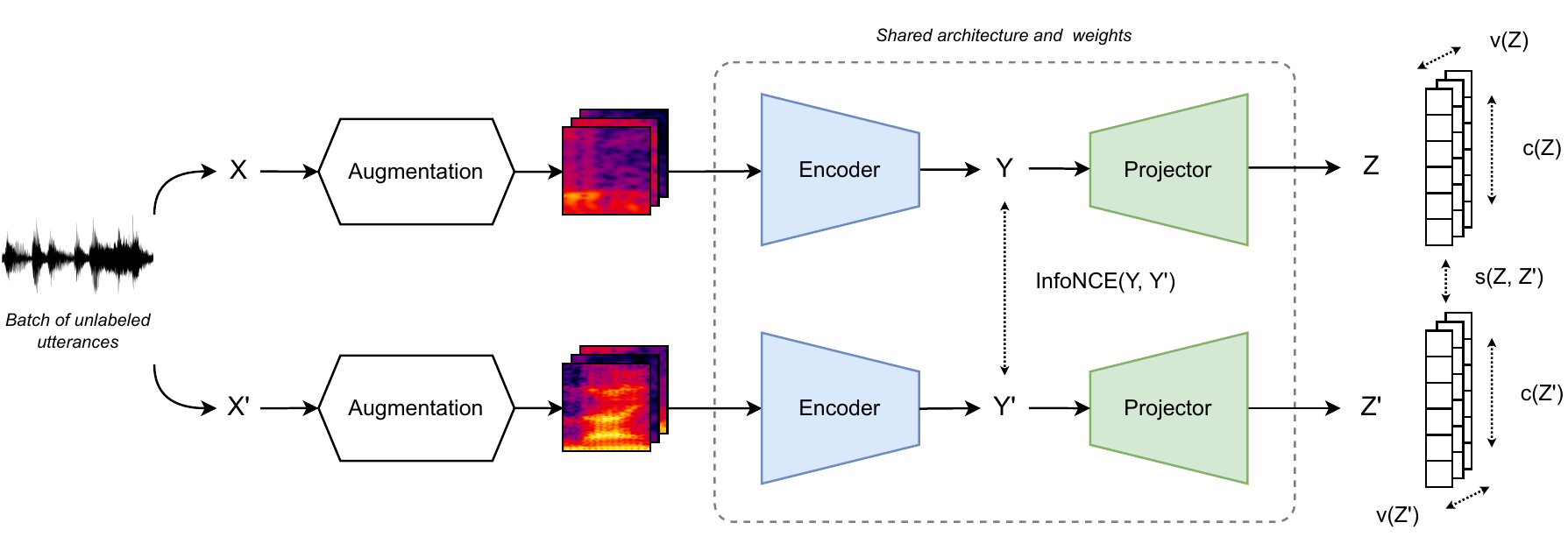}
  \caption{Diagram of our self-supervised training framework.}
  \label{fig:training_framework}
\end{figure*}

To assess the ability of these non-contrastive methods to be effective on the task of speaker verification, we develop a self-supervised training framework using an \textit{information maximization} objective function to learn speaker representations from unlabeled data. Our training framework is further described in Section 2. In Section 3, we present our experimental setup and our intensive data augmentation step as state-of-the-art methods \cite{Snyder2018Xvectors, Zhang2021SimCLR, Xia2021MOCOSV} have proven that it is fundamental to produce channel invariant speaker embeddings. We compare the  performances of our model based on our \textit{information maximization} loss with a traditional contrastive loss in Section 4. Furthermore, we evaluate the ability of both objective functions to be complementary when trained jointly. Through a label-efficient evaluation, we show that our approach can reach the performance of the supervised baseline with a limited amount of labeled utterances. Finally, we conclude in Section 5.

\section{Methods}
\label{sec:2}


Our self-supervised training framework is depicted in Figure \ref{fig:training_framework}. The architecture is based on a simple Siamese neural network to produce a pair of embeddings for a given unlabeled utterance.  For each mini-batch, we randomly sample $N$ utterances from the dataset. We extract two non-overlapping frames, denoted as $\mathbf{x}$ and $\mathbf{x'}$, for each utterance. Then, we apply different random augmentations to both copies and use their mel-scaled spectrogram as features for the neural network. As only speaker identity must be extracted, data augmentation is fundamental to avoid encoding channel information into the representations.
An encoder transforms $\mathbf{x}$ and $\mathbf{x'}$ to their respective representations $\mathbf{y}$ and $\mathbf{y'}$. Then, the representations are fed to a projector to compute the embeddings $\mathbf{z}$ and $\mathbf{z'}$. Thus, for each mini-batch, we stack $\mathbf{x}$ samples into $\mathbf{X}$ and $\mathbf{x'}$ samples into $\mathbf{X'}$ to produce representations ($\mathbf{Y}$, $\mathbf{Y'}$) and embeddings ($\mathbf{Z}$, $\mathbf{Z'}$). The model architecture is described in further detail in the following section. Representations are used to perform speaker verification while embeddings are used only during training. Different objective functions are explored to optimize the model.

\subsection{Contrastive learning}

Contrastive learning aims to maximize the similarity between positive pairs while minimizing the similarity between negative pairs. Positive pairs are constructed with embeddings $\mathbf{z}_i$ and $\mathbf{z}_i^{\prime}$ that are derived from the same utterance and thus share the same speaker identity. Negative pairs are created using $\mathbf{z}_i$ and all other embeddings in the mini-batch $\mathbf{z}_j$ ($1 \leq j \leq N$). We rely on a loss function based on InfoNCE \cite{Oord2019Representation} defined as:
\begin{align}
    \mathcal{L}_{\mathrm{InfoNCE}}=\frac{1}{N} \sum_{i=1}^{N}-\log \frac{\exp \left(\mathbf{z}_{i} \cdot \mathbf{z}_{i}^{\prime} / \tau\right)}{\sum_{j=1}^{N} \exp \left(\mathbf{z}_{i} \cdot \mathbf{z}_{j} / \tau\right)}
\end{align}
where $\tau$ is a temperature-scaling hyper-parameter. As we normalize speaker embeddings to unit vectors ($\left\|\mathbf{z}_i\right\|=1$), their dot product is equal to their cosine similarity.

\subsection{Information maximization}

\subsubsection{Barlow Twins}

Barlow Twins \cite{Zbontar2021Barlow} objective function does not require contrastive samples and maximize information directly from the embeddings as it operates on the cross-correlation matrix $\mathcal{C}$ between $\mathbf{Z}$ and $\mathbf{Z}^{\prime}$. The objective is to make $\mathcal{C}$ similar to the identity matrix. The \textit{invariance term}, by making the on-diagonal coefficients to be 1, forces the embeddings to be invariant to the distortions applied. The \textit{redundancy reduction term}, by pushing all coefficients off-diagonal to be $0$, decorrelates the different vector components and thus reduces the redundancy between them. The loss is defined as:
\begin{multline}
    \mathcal{L}_{\rm BarlowTwins}
    = \sum_{i}\left(1-\left[\mathcal{C}\left(\mathbf{Z}, \mathbf{Z}^{\prime}\right)\right]_{i i}\right)^{2} \\ 
    + \lambda \sum_{i} \sum_{j \neq i} \left[\mathcal{C}\left(\mathbf{Z}, \mathbf{Z}^{\prime}\right)\right]_{i j}^{2}
\end{multline}
where $\lambda$ is a hyper-parameter to scale the redundancy reduction term.

\subsubsection{Variance-Invariance-Covariance Regularization}

VICReg \cite{Bardes2021VICReg} is based on three training objectives and provides a simple and explicit solution to avoid collapse. The complete objective function is defined as:
\begin{multline}
    \label{eq:vicreg}
    \mathcal{L}_{\mathrm{VICReg}}
    = \lambda \, s\left(\mathbf{Z}, \mathbf{Z}^{\prime}\right) 
    + \mu\left(v(\mathbf{Z})+v\left(\mathbf{Z}^{\prime}\right)\right) \\
    + \nu\left(c(\mathbf{Z})+c\left(\mathbf{Z}^{\prime}\right)\right) 
\end{multline}
where $\lambda$, $\mu$ and $\nu$ are hyper-parameters to scale the \textit{variance}, \textit{invariance} and \textit{covariance} terms. $s$, $v$ and $c$ represent the invariance, variance and covariance components, respectively.

\paragraph{Variance} As we assume that each sample in the mini-batch is from a unique speaker, we enforce the variance to reach $1$ along the $D$ dimensions of the embeddings. The objective is to produce unique representations for each speaker and prevent a collapsing solution.
\begin{align}
    \label{eq:v}
    v\left(\mathbf{Z}\right)=\frac{1}{D} \sum_{j=1}^{D} \max \left(0, 1-\sqrt{\operatorname{Var}(\mathbf{z}^j)}\right)
\end{align}
\paragraph{Invariance} By reducing the $l_2$ distance between $\mathbf{z}_{i}$ and $\mathbf{z}_{i}^{\prime}$, which have been augmented differently, we learn invariance to multiple views of the same data (noise, reverberation, etc.).
\begin{align}
    s\left(\mathbf{Z}, \mathbf{Z}^{\prime}\right)=\frac{1}{N} \sum_{i=1}^{N}\left\|\mathbf{z}_{i}-\mathbf{z}_{i}^{\prime}\right\|_{2}^{2}
\end{align}

\paragraph{Covariance} To decorrelate the different dimensions of the projections, we make the off-diagonal coefficients of the covariance matrix $C$ to be close to $0$. The underlying intuition is to prevent dimensions from encoding similar information.
\begin{align}
    c\left(\mathbf{Z}\right)=\frac{1}{D} \sum_{i \neq j}[C(\mathbf{Z})]_{i, j}^{2}
\end{align}

\subsection{Exploring the complementarity of these methods}
\label{sec:2.4}

We explore the ability of both methods to be complementary by training our model jointly with information maximization and contrastive learning techniques. For this purpose, we optimize our framework using  $\mathcal{L}_{\mathrm{comp}}^{\mathrm{1}}$ and $\mathcal{L}_{\mathrm{comp}}^{\mathrm{2}}$. It is noteworthy that their respective components operate at different stages of the neural network, either on representations or on embeddings.
\begin{align}
    \mathcal{L}_{\mathrm{comp}}^{\mathrm{1}}
    &= \mathcal{L}_{\mathrm{VICReg}} \left(  \mathbf{Y}, \mathbf{Y}^{\prime} \right) + \mathcal{L}_{\mathrm{InfoNCE}} \left(  \mathbf{Z}, \mathbf{Z}^{\prime} \right) \label{eqn:L1comp}\\
    \mathcal{L}_{\mathrm{comp}}^{\mathrm{2}}
    &= \mathcal{L}_{\mathrm{InfoNCE}} \left(  \mathbf{Y}, \mathbf{Y}^{\prime} \right) + \mathcal{L}_{\mathrm{VICReg}} \left(  \mathbf{Z}, \mathbf{Z}^{\prime} \right) \label{eqn:L2comp}
\end{align}

Moreover, we rely on $\mathcal{L}_{\mathrm{reg}}^{\mathrm{Y}}$ and $\mathcal{L}_{\mathrm{reg}}^{\mathrm{Z}}$ to determine if information maximization methods can be used as an interesting regulariation term for the contrastive learning objective function. $\alpha$ is set to $0.1$ in our experiments.
\begin{align}
    \mathcal{L}_{\mathrm{reg}}^{\mathrm{Y}}
    &= \mathcal{L}_{\mathrm{InfoNCE}} \left(  \mathbf{Y}, \mathbf{Y}^{\prime} \right) + \alpha \, \mathcal{L}_{\mathrm{VICReg}} \left(\mathbf{Y},\mathbf{Y}^{\prime}\right) \\
    \mathcal{L}_{\mathrm{reg}}^{\mathrm{Z}}
    &= \mathcal{L}_{\mathrm{InfoNCE}} \left(  \mathbf{Z}, \mathbf{Z}^{\prime} \right) + \alpha \, \mathcal{L}_{\mathrm{VICReg}} \left(  \mathbf{Z}, \mathbf{Z}^{\prime} \right) \label{eqn:LregZ}
\end{align}
\section{Experimental setup}

\subsection{Datasets and feature extraction}

Experiments are performed on the VoxCeleb1 \cite{Nagrani2017VoxCeleb} dataset that contains 148,642 utterances from 1,211 speakers. The proposed framework is trained on the \textit{dev} set and is evaluated on the \textit{test} set which is composed of 4,874 utterances from 40 speakers. Speaker labels are discarded during self-supervised training.
We use audio chunks of 2 seconds per sample and extract 40-dimensional log-mel spectrogram features with a Hamming window of length 25 ms and a 10 ms frame-shift. Speech activity detection (SAD) is not performed as training data consists mostly of continuous speech. Mean and variance normalization (MVN) is applied using instance normalization on the network input features.

\subsection{Data augmentation}

Self-supervised learning frameworks tend to rely on extensive augmentation techniques to produce representations robust against extrinsic variabilities. For speaker verification, we want to learn embeddings invariant to channel information such as noise coming from the environment or the recording device. The segments extracted from the same audio files share the same channel and noise background. Thus, if we want speaker identity to be the only differentiating factor between two representations, it is essential to provide different views of the same utterance to avoid encoding channel characteristics. During training, we apply transformations to the input signal randomly at each training step. Utterances are first augmented with background noises, overlapping music tracks or speech segments before being reverberated. We rely on the MUSAN corpus \cite{Snyder2015Musan} which consists of about 60 hours of speech, 42h of music from different genres and a total of 929 various noises. The SNR is uniformly sampled between $13$ and $20$ dB for speech, $5$ and $15$ dB for music and $0$ and $15$ dB for noises. To apply reverberation, we convolve the input audio segment with an impulse response randomly sampled from the Simulated Room Impulse Response Database \cite{Ko2017RIR}.

\subsection{Model architecture and training}

The encoder is based on the Thin-ResNet34 \cite{Chung2020Delving} architecture with 1024 output units. Self-attentive pooling (SAP) \cite{Cai2018SAP} is employed to produce utterance-level representations. The projector is a standard MLP composed of three fully-connected layers of 2048 units. The two intermediate layers are followed by Batch Normalization (BN) and ReLU non-linearity. For InfoNCE loss, we compute the dot product of $l_2$-normalized vector and use $\tau=0.07$ as the temperature hyper-parameter. By default, we use $\lambda=1$, $\mu=1$ and $\nu=0.04$ for VICReg and $\lambda=0.05$ for Barlow Twins.
We use Adam optimizer with a learning rate of $0.001$ reduced by $5\%$ every 10 epochs. We train for 500 epochs with a default batch size of 256 and we stop the training if the test EER is not improving after 50 epochs. Our implementation is based on PyTorch and we run our experiments on 2 NVIDIA Titan X (Pascal) 12 GB GPUs.

\subsection{Evaluation protocol}

For each utterance in the test set, we extract the representations from a fixed number of evenly spaced frames before averaging them across the temporal axis. The scoring is determined using the cosine similarity between two $l_2$-normalized representations. The performance is reported in terms of Equal Error Rate (EER) and minimum Detection Cost Function (minDCF). Based on NIST 2016 Speaker Recognition Evaluation protocol, we use $P_{\rm target} = 0.01$ for the minDCF.
\section{Results and discussions}

\subsection{The impact of data augmentation}

The performance of our model when trained with different data augmentation strategies is shown in Table \ref{tab:data-augmentation}. Our baseline trained without applying transformations on the input achieves 29.87\% EER. The results are drastically improved when adding background noises and reverberation, reaching 21.22\% EER and 22.28\% EER, respectively. The best performance is achieved when applying both methods simultaneously as it results in 11.14\% EER and 0.6843 minDCF which represent a 62.7\% relative improvement of the EER. Therefore, data augmentation is fundamental when training a self-supervised speaker verification framework.

\begin{table}[h]
  \caption{The performance of our self-supervised SV system when trained with different data augmentation strategies.}
  \label{tab:data-augmentation}
  \centering
  \begin{tabular}{lccc}
    \toprule
    \textbf{Method}   & \textbf{EER}  & \textbf{minDCF}  \\
    \midrule
    No augmentation   & 29.87            & 0.8833           \\
    Musan             & 21.22            & 0.8388           \\
    RIR               & 22.28            & 0.8525           \\
    Musan + RIR       & \textbf{11.14} & \textbf{0.6843}  \\
    \bottomrule
  \end{tabular}
\end{table}

\subsection{Evaluating the role of VICReg components}

Table \ref{tab:vicreg-coefficients} reports the performance for various values of VICReg loss coefficients from (Eq. \ref{eq:vicreg}). The weight of the invariance term is fixed as it is fundamental to learn representations invariant to different transformations and to benefit from data augmentation. Moreover, we have determined that the variance regularization term is absolutely necessary to avoid collapse and that other terms must be scaled accordingly to its weight. The covariance objective has also a significant impact on the results as we observed a 53.6\% improvement of the EER between the first and third configuration. Finally, we use the same hyper-parameters as the original work as it produces the best performance by achieving 11.14\% EER and 0.6843 minDCF.

\begin{table}[t]
  \caption{The impact of different scaling factors for VICReg loss components: $\lambda$ (Invariance), $\mu$ (Variance) and $\nu$ (Covariance).}
  \label{tab:vicreg-coefficients}
  \centering
  \begin{tabular}{cccccc}
    \toprule
    \textbf{$\lambda$}  & \textbf{$\mu$}  & \textbf{$\nu$}  & \textbf{EER}  & \textbf{minDCF}  \\
    \midrule
    1                   & 1               & 0               & 24.00            & 0.9964          \\
    1                   & 0.5             & 0.1             & 15.71            & 0.8554           \\
    1                   & 1               & 0.04            & \textbf{11.14}   & \textbf{0.6843}           \\
    1                   & 1               & 0.1             & 11.87            & 0.7101        \\
    \bottomrule
  \end{tabular}
\end{table}

\subsection{Effect of projector dimensionality}

Similarly to many self-supervised methods, we observe a dependency between the dimensionality of the projector network and the performance on the downstream task. As shown in Table~\ref{tab:projector-dim}, the EER dramatically decreases from 14.96\% without a projector to 10.77\% with a projector of 1024 hidden and output units. This suggests that self-supervised learning benefits from large projector architectures. We hypothesize that the covariance mechanism benefits from a larger dimensionality to spread the information more efficiently.

\begin{table}[t]
  \caption{Effect of projector dimensionality (number of hidden and output units) on the performance of our self-supervised SV system.}
  \label{tab:projector-dim}
  \centering
  \begin{tabular}{lccc}
    \toprule
    \textbf{Architecture}  & \textbf{EER}  & \textbf{minDCF}  \\
    \midrule
    No projector           & 14.96            & 0.9369           \\
    512                    & 11.34            & 0.7826           \\
    1024                   & \textbf{10.77}   & 0.7208           \\
    2048                   & 11.14            & \textbf{0.6843}           \\
    \bottomrule
  \end{tabular}
\end{table}

\begin{table}[h!]
  \caption{Self-supervised SV results on VoxCeleb1 test set.}
  \label{tab:results}
  \centering
  \begin{tabular}{llccc}
    \toprule    \textbf{Method} & \textbf{Loss}  & \textbf{EER}  & \textbf{minDCF}  \\
    \midrule
    NPC \cite{Jati2019NPC} & Cross-entropy & 15.54 & 0.8700 \\
    SimCLR \cite{Zhang2021SimCLR} & InfoNCE & 9.87 & 0.6760 \\
    \midrule
     & $\mathcal{L}_{\mathrm{InfoNCE}}$       & 10.42            & \textbf{0.6276}           \\
    Ours & $\mathcal{L}_{\mathrm{BarlowTwins}}$  & 13.46            & 0.8473           \\
     & $\mathcal{L}_{\mathrm{VICReg}}$        & \textbf{9.25}    & 0.6432           \\
     \midrule
                & $\mathcal{L}_{\mathrm{comp}}^{\mathrm{1}}$   &         13.14  &         0.6950   \\
     Ours       & $\mathcal{L}_{\mathrm{comp}}^{\mathrm{2}}$   &  \textbf{8.47} & \textbf{0.6400}   \\
     (Section \ref{sec:2.4}) & $\mathcal{L}_{\mathrm{reg}}^{\mathrm{Y}}$   &          9.09  &         0.6894   \\
                & $\mathcal{L}_{\mathrm{reg}}^{\mathrm{Z}}$   &         10.38  & 0.6913  \\
    \bottomrule
  \end{tabular}
\end{table}




\subsection{Complementarity of information maximization and contrastive learning methods}

Table \ref{tab:results} reports downstream results of the different methods described in Section \ref{sec:2} and some existing unsupervised techniques. Our method based on VICReg achieves 9.25\% EER and outperforms InfoNCE (10.42\% EER), BarlowTwins (13.46\% EER) and the other unsupervised approaches on the same corpus. Even though the best minDCF is obtained with InfoNCE, competitive results can be obtained without relying on contrastive samples. Moreover, we reach the best performance with $\mathcal{L}_{\mathrm{comp}}^{\mathrm{2}}$ (Eq. \ref{eqn:L2comp}) objective function as it achieves 8.47\% EER. Therefore, combining information maximization and contrastive learning, each at a different stage of the model, is very effective. As a result, the learned speaker representations are more accurate implying that the two approaches are complementary.

\begin{figure}[h!]
    \centering
    \includegraphics[width=\linewidth]{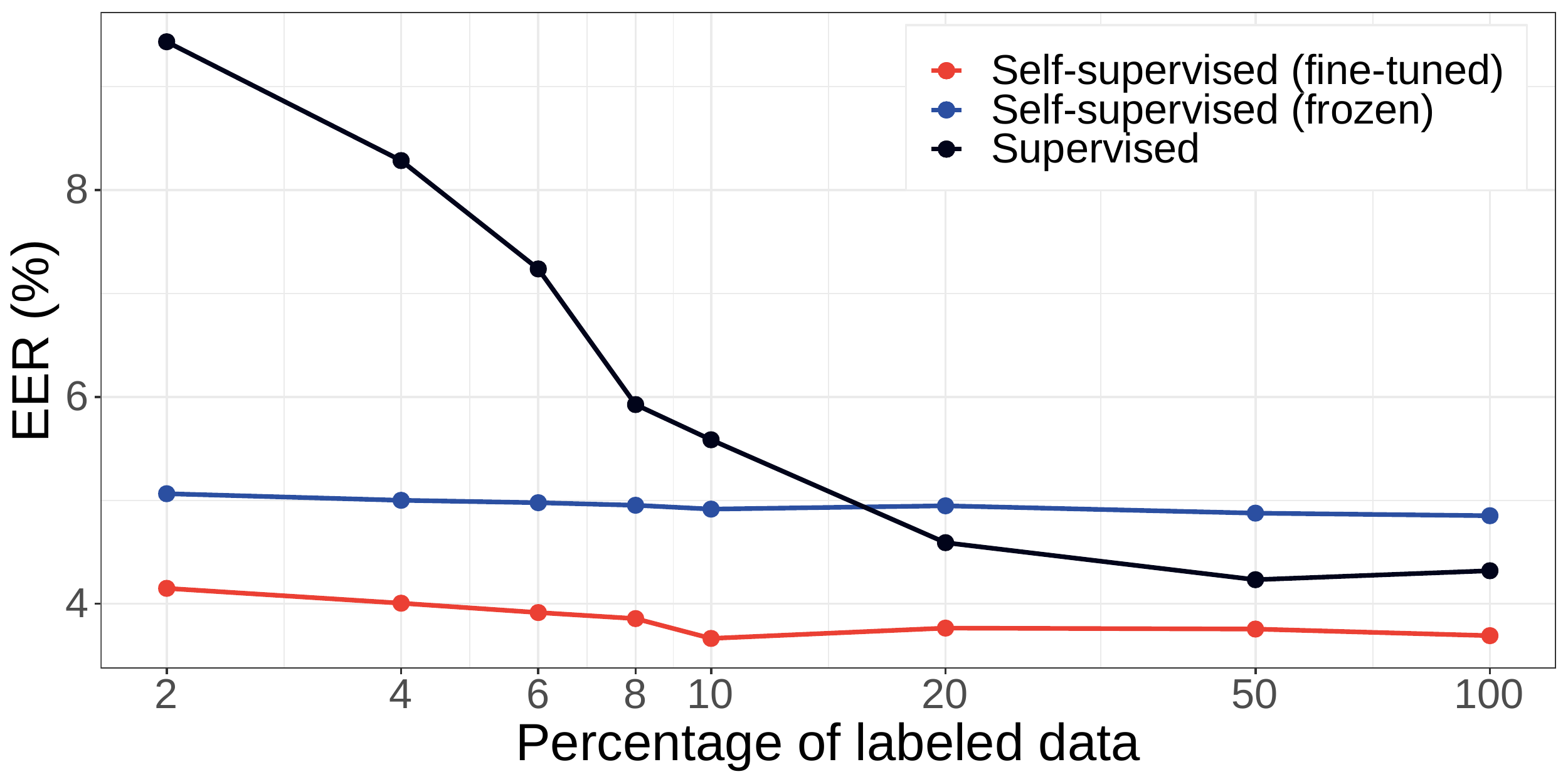}
    \caption{Results on SV with different percentage of labeled data used during training.}
    \label{fig:label_efficient}
\end{figure}
\subsection{Label-efficient evaluation}

We explore two ways to improve the performance of our self-supervised framework with a limited amount of annotated utterances: (1) train a linear classifier on top of the frozen self-supervised representations ; (2) fine-tune the whole pre-trained neural network. As shown in Figure \ref{fig:label_efficient}, decreasing the amount of labeled data has almost no impact on the proposed self-supervised method. Furthermore, fine-tuning with only 2\% of labeled data is sufficient to outperform the supervised baseline. Even when using all labels, we achieve 3.38\% EER with fine-tuning compared to 4.3\% EER for the supervised counterpart.
\section{Conclusion}

In this study, we propose a self-supervised training framework based on information maximization techniques to learn robust speaker representations without contrastive samples. We demonstrate the effectiveness of this method which achieves better performance compared to existing self-supervised techniques on speaker verification. Furthermore, we show its ability to complement a contrastive objective function, operating on the representations, by maximizing the information in the embeddings. Finally, our method outperforms its supervised counterpart when fine-tuned with only 2\% of labeled utterances, which is a step toward label-efficient speaker verification systems.

\def\url#1{}
\bibliographystyle{IEEEtran}
\bibliography{ssl_for_sv}

\begin{thebibliography}{10}
\providecommand{\url}[1]{#1}
\csname url@samestyle\endcsname
\providecommand{\newblock}{\relax}
\providecommand{\bibinfo}[2]{#2}
\providecommand{\BIBentrySTDinterwordspacing}{\spaceskip=0pt\relax}
\providecommand{\BIBentryALTinterwordstretchfactor}{4}
\providecommand{\BIBentryALTinterwordspacing}{\spaceskip=\fontdimen2\font plus
\BIBentryALTinterwordstretchfactor\fontdimen3\font minus
  \fontdimen4\font\relax}
\providecommand{\BIBforeignlanguage}[2]{{%
\expandafter\ifx\csname l@#1\endcsname\relax
\typeout{** WARNING: IEEEtran.bst: No hyphenation pattern has been}%
\typeout{** loaded for the language `#1'. Using the pattern for}%
\typeout{** the default language instead.}%
\else
\language=\csname l@#1\endcsname
\fi
#2}}
\providecommand{\BIBdecl}{\relax}
\BIBdecl

\bibitem{dehakivector2011}
N.~Dehak, P.~J. Kenny, R.~Dehak, P.~Dumouchel, and P.~Ouellet, ``Front-end
  factor analysis for speaker verification,'' \emph{{IEEE} Transactions on
  Audio, Speech, and Language Processing}, vol.~19, no.~4, pp. 788--798, 2011.

\bibitem{Snyder2018Xvectors}
D.~Snyder, D.~Garcia-Romero, G.~Sell, D.~Povey, and S.~Khudanpur, ``{X-Vectors:
  Robust DNN Embeddings for Speaker Recognition},'' in \emph{ICASSP}, 2018, pp.
  5329--5333.

\bibitem{villalba2020}
J.~Villalba, N.~Chen, D.~Snyder, D.~Garcia-Romero, A.~{McCree}, G.~Sell,
  J.~Borgstrom, L.~P. García-Perera, F.~Richardson, R.~Dehak, P.~A.
  Torres-Carrasquillo, and N.~Dehak, ``State-of-the-art speaker recognition
  with neural network embeddings in {NIST} {SRE}18 and speakers in the wild
  evaluations,'' \emph{Computer Speech \& Language}, vol.~60, p. 101026, 2020.

\bibitem{Chung2020Delving}
J.~S. Chung, J.~Huh, and S.~Mun, ``{Delving into VoxCeleb: Environment
  Invariant Speaker Recognition},'' in \emph{Odyssey}, 2020, pp. 349--356.

\bibitem{Chung2020Metric}
J.~S. Chung, J.~Huh, S.~Mun, M.~Lee, H.-S. Heo, S.~Choe, C.~Ham, S.~Jung, B.-J.
  Lee, and I.~Han, ``{In Defence of Metric Learning for Speaker Recognition},''
  in \emph{Interspeech}, 2020, pp. 2977--2981.

\bibitem{Zhang2021SimCLR}
H.~Zhang, Y.~Zou, and H.~Wang, ``{Contrastive Self-Supervised Learning for
  Text-Independent Speaker Verification},'' in \emph{ICASSP}, 2021, pp.
  6713--6717.

\bibitem{Xia2021MOCOSV}
W.~Xia, C.~Zhang, C.~Weng, M.~Yu, and D.~Yu, ``{Self-supervised
  Text-independent Speaker Verification using Prototypical Momentum Contrastive
  Learning},'' in \emph{ICASSP}, 2021, pp. 6723--6727.

\bibitem{Oord2019Representation}
A.~van~den Oord, Y.~Li, and O.~Vinyals, ``{Representation Learning with
  Contrastive Predictive Coding},'' \emph{arXiv preprint arXiv:1807.03748},
  2019.

\bibitem{Chen2020SimCLR}
\BIBentryALTinterwordspacing
T.~Chen, S.~Kornblith, M.~Norouzi, and G.~Hinton, ``{A Simple Framework for
  Contrastive Learning of Visual Representations},'' in \emph{ICML}, 2020, pp.
  1597--1607.  \url{https://proceedings.mlr.press/v119/chen20j.html}
\BIBentrySTDinterwordspacing

\bibitem{He2020MOCO}
K.~He, H.~Fan, Y.~Wu, S.~Xie, and R.~Girshick, ``{Momentum Contrast for
  Unsupervised Visual Representation Learning},'' in \emph{CVPR}, 2020, pp.
  9726--9735.

\bibitem{chen2020SimSiam}
X.~Chen and K.~He, ``{Exploring Simple Siamese Representation Learning},'' in
  \emph{CVPR}, 2021, pp. 15\,745--15\,753.

\bibitem{Grill2020BYOL}
\BIBentryALTinterwordspacing
J.-B. Grill, F.~Strub, F.~Altché, C.~Tallec, P.~H. Richemond, E.~Buchatskaya,
  C.~Doersch, B.~A. Pires, Z.~D. Guo, M.~G. Azar, B.~Piot, K.~Kavukcuoglu,
  R.~Munos, and M.~Valko, ``{Bootstrap your own latent: A new approach to
  self-supervised Learning},'' in \emph{NeurIPS}, 2020, pp. 21\,271--21\,284.
  \url{https://proceedings.neurips.cc/paper/2020/hash/f3ada80d5c4ee70142b17b8192b2958e-Abstract.html}
\BIBentrySTDinterwordspacing

\bibitem{Caron2021SwAV}
\BIBentryALTinterwordspacing
M.~Caron, I.~Misra, J.~Mairal, P.~Goyal, P.~Bojanowski, and A.~Joulin,
  ``{Unsupervised Learning of Visual Features by Contrasting Cluster
  Assignments},'' in \emph{NeurIPS}, 2021, pp. 9912--9924.
  \url{https://proceedings.neurips.cc/paper/2020/hash/70feb62b69f16e0238f741fab228fec2-Abstract.html}
\BIBentrySTDinterwordspacing

\bibitem{Caron2018DeepCluster}
M.~Caron, P.~Bojanowski, A.~Joulin, and M.~Douze, ``{Deep Clustering for
  Unsupervised Learning of Visual Features},'' in \emph{ECCV}, 2018, pp.
  139--156.

\bibitem{Bardes2021VICReg}
A.~Bardes, J.~Ponce, and Y.~LeCun, ``{VICReg: Variance-Invariance-Covariance
  Regularization for Self-Supervised Learning},'' \emph{arXiv preprint
  arXiv:2105.04906}, 2021.

\bibitem{Zbontar2021Barlow}
\BIBentryALTinterwordspacing
J.~Zbontar, L.~Jing, I.~Misra, Y.~LeCun, and S.~Deny, ``{Barlow Twins:
  Self-Supervised Learning via Redundancy Reduction},'' in \emph{ICML}, 2021,
  pp. 12\,310--12\,320.
  \url{https://proceedings.mlr.press/v139/zbontar21a.html}
\BIBentrySTDinterwordspacing

\bibitem{Nagrani2017VoxCeleb}
A.~Nagrani, J.~S. Chung, and A.~Zisserman, ``{VoxCeleb: A Large-Scale Speaker
  Identification Dataset},'' in \emph{Interspeech}, 2017, pp. 2616--2620.

\bibitem{Snyder2015Musan}
D.~Snyder, G.~Chen, and D.~Povey, ``{MUSAN: A Music, Speech, and Noise
  Corpus},'' \emph{arXiv preprint arXiv:1510.08484}, 2015.

\bibitem{Ko2017RIR}
T.~Ko, V.~Peddinti, D.~Povey, M.~L. Seltzer, and S.~Khudanpur, ``{A study on
  data augmentation of reverberant speech for robust speech recognition},'' in
  \emph{ICASSP}, 2017, pp. 5220--5224.

\bibitem{Cai2018SAP}
W.~Cai, J.~Chen, and M.~Li, ``{Exploring the Encoding Layer and Loss Function
  in End-to-End Speaker and Language Recognition System},'' in \emph{Odyssey},
  2018.

\bibitem{Jati2019NPC}
\BIBentryALTinterwordspacing
A.~Jati and P.~Georgiou, ``Neural predictive coding using convolutional neural
  networks toward unsupervised learning of speaker characteristics,'' in
  \emph{IEEE/ACM Transactions on Audio, Speech, and Language Processing}, 2019,
  p. 1577–1589.  \url{http://dx.doi.org/10.1109/TASLP.2019.2921890}
\BIBentrySTDinterwordspacing

\end{thebibliography}

\end{document}